\documentstyle[aps,prl,epsf,twocolumn,floats]{revtex}

\begin{document}
\draft
\title{Fluctuating spin $g$-tensor in small metal grains}

\author{P. W. Brouwer$^{a}$, X. Waintal$^{a}$, and B. I. Halperin$^{b}$}
\address{$^{a}$Laboratory of Atomic and Solid State Physics,
Cornell University, Ithaca, NY 14853-2501\\
$^{b}$Lyman Laboratory of Physics, Harvard University, Cambridge MA 
02138\\
{\rm \today}
\medskip \\ \parbox{14cm}{\rm
In the presence of spin-orbit scattering, the splitting of an energy
level $\varepsilon_{\mu}$ in a generic small metal grain due to the
Zeeman coupling to a magnetic field $\vec B$ depends on the direction
of $\vec B$, as a result of mesoscopic fluctuations. The anisotropy is
described by the eigenvalues $g_j^2$ ($j=1,2,3$) of a tensor ${\cal
G}$, corresponding to the (squares of) $g$-factors along three
principal axes. We consider the statistical distribution of ${\cal G}$
and find that the anisotropy is enhanced by eigenvalue repulsion between
the $g_{j}$.
}}

% 71. Electronic structure (see also 73.20 Surface 
%     and interface electron states)
% 71.24.+q Electronic structure of clusters and nanoparticles
% 71.70.Ej Spin-orbit coupling, Zeeman and Stark splitting, 
% Jahn-Teller effect

\maketitle

With the advance of nanoparticle technology, it has become possible to
resolve individual energy levels for electrons in ultrasmall metal
grains. Recent experiments addressed their Zeeman
splitting under the application of a magnetic field $\vec B$
\cite{Ralph,Davidovic,Salinas}. The splitting of a level
$\varepsilon_{\mu}$ is described by a $g$-factor, $\delta
\varepsilon_{\mu} = \pm \case{1}{2} \mu_B g B_z$, where $\mu_B$ is the Bohr
magneton. A free electron has $g = 2$, but in
small metal grains the effective $g$-factor may be reduced as a result
of spin-orbit scattering \cite{Halperin}. 
In order to study this reduction, Salinas
{\rm et al.} \cite{Salinas} have doped Al grains (which do not have
significant spin-orbit scattering) with Au (which has). For small
concentrations of Au, the effective $g$-factor was seen to drop
from 2 to around 0.7. Even lower values $g \sim 0.3$ were reported in 
experiments on Au grains \cite{Davidovic}.

For disordered systems with spin-orbit scattering, the splitting of
a level $\varepsilon_{\mu}$ does not only depend on the magnitude of
the magnetic field $\vec B$, but also on its direction. Hence, an
analysis in terms of a ``$g$-tensor'' is more appropriate
\cite{Slichter}. To be precise, 
the Zeeman field splits the Kramers' doublet $\varepsilon_{\mu} \to
\varepsilon_{\mu} \pm \delta \varepsilon_{\mu}$ with
\begin{eqnarray}
\delta \varepsilon_{\mu}^2 = (\mu_B/2)^2 \vec B \cdot {\cal G_{\mu}} 
  \cdot \vec B,
  \label{eq:deltaE}
\end{eqnarray}
where ${\cal G}_{\mu}$ is a $3 \times 3$ tensor.  In the absence of
spin-orbit scattering, the tensor ${\cal G}_{\mu}$ is isotropic,
$({\cal G_{\mu}})_{ij} = 4 \delta_{ij}$. The effect of spin-orbit
scattering on ${\cal G}_{\mu}$ is threefold: It
leads to a decrease of the typical magnitude of ${\cal G}_{\mu}$, it
makes the tensor structure of ${\cal G}_{\mu}$ important (i.e., it
introduces an anisotropic response to the magnetic field $\vec B$),
and it causes ${\cal G}_{\mu}$ to be different for each level
$\varepsilon_{\mu}$. Hence ${\cal G}_{\mu}$ becomes a fluctuating
quantity, and it is
important to know its statistical distribution.  The latter problem
was addressed in a recent paper by Matveev et al.\ \cite{Matveev},
however without considering the tensor structure of ${\cal
G}_{\mu}$. The anisotropy of the $g$-tensor is a measurable quantity and
we here consider the distribution of the entire tensor ${\cal G}_{\mu}$.
The distribution $P({\cal G}_{\mu})$ is defined with respect to an ensemble 
of small metal grains of roughly equal size. The same distribution applies
to the fluctuations of ${\cal G}_{\mu}$ as a function of the level 
$\varepsilon_{\mu}$ in the same grain.

In general, ${\cal G}_{\mu}$ has a contribution ${\cal
G}_{\mu}^{\rm spin}$ from the magnetic moment of electron spins, and a
contribution ${\cal G}_{\mu}^{\rm orb}$ for the orbital angular moment of
the state $|\psi_{\mu}\rangle$. In Ref.\ \onlinecite{Matveev}, the
typical sizes of both contributions were estimated as ${\cal G}^{\rm
spin} \sim \tau_{\rm so} \Delta$ and ${\cal G}^{\rm orb} \sim \ell/L$,
where $\tau_{\rm so}$ is the mean spin-orbit scattering time, $L$ is
the grain size, $\Delta \propto L^{-3}$ is the mean level spacing, and
$\ell \ll L$ is the elastic mean free path.
We restrict ourselves to the spin
contribution ${\cal G}^{\rm spin}$, which should be dominant for small
grain sizes \cite{Matveev}, provided $\tau_{\rm so}$ does not
depend on system size, as should be the case for the experiments of
Ref.\ \onlinecite{Salinas}. 
When orbital contributions are important, the anisotropy of ${\cal G}$
will be affected by the shape of the grain. In that case, our main
conclusions apply only to a roughly spherical grain.
As the typical magnitude of ${\cal G}$
(we drop the superscript ``spin'' and the subscript $\mu$ if there
is no ambiguity)
depends on the microscopic parameters $\tau_{\rm so}$ and $\Delta$,
which are in most cases not known accurately, we choose to have
the typical magnitude of ${\cal G}$ serve as an external parameter in
our theory. 

\begin{figure}
\vglue -0.5cm
\epsfxsize=0.99\hsize
%\hspace{0.1\hsize}
\epsffile{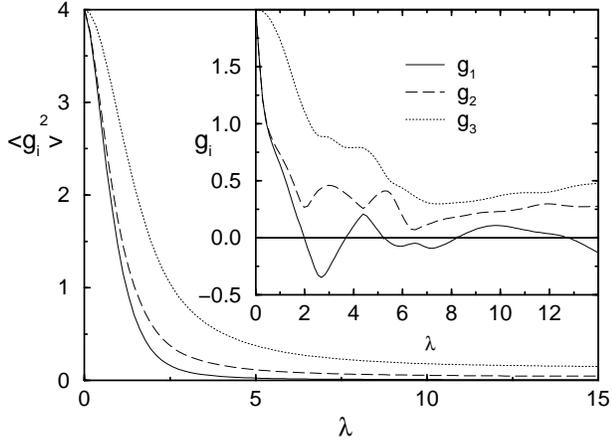}

\caption{\label{fig:3} Average of the squares of principal 
$g$-factors versus spin-orbit scattering strength
$\lambda$, obtained from numerical simulation of the random matrix
model (\protect\ref{eq:HSA}) with $N=100$. Inset: 
$g_1$, $g_2$, and $g_3$ for a
specific realization. We have included the sign of $g_1$; see the
discussion below Eq.\ (\protect\ref{eq:ga}).}
\end{figure}

We first present our main results. 
With a suitable choice of the coordinate axes (``principal
axes''), the tensor ${\cal G}$ can be diagonalized. Writing its
eigenvalues as $g_j^2$ and denoting the components of the magnetic
field along the principal axes by $B_j$, $j=1,2,3$, Eq.\ (\ref{eq:deltaE})
takes a particularly simple form,
\begin{equation}
  \delta \varepsilon_{\mu}^2 = \case{1}{4} \mu_B^2 (
    g_1^2 B_1^2 + g_2^2 B_2^2 + g_3^3 B_3^2). \label{eq:deltaE2}
\end{equation}
We refer to the numbers $g_1$, $g_2$, and $g_3$ as principal
$g$-factors.  For a generic metal grain of a cubic material, 
rotational
symmetry implies that, for a given level $\varepsilon_{\mu}$, the
positioning of the principal axes is entirely random in space,
as long as they are mutually orthogonal. Hence, it remains to
study the distribution $P(g_1,g_2,g_3)$ of the principal $g$-factors
$g_1$, $g_2$, and $g_3$ for the level $\varepsilon_{\mu}$. Our main
result is, that for sufficiently strong spin-orbit scattering,
$P(g_1,g_2,g_3)$ is given by the distribution
\begin{equation}
  P(g_1,g_2,g_3) \propto \prod_{i<j} |g_i^2 - g_j^2| \prod_{i} 
  e^{-3 g_i^2/2 \langle g^2 \rangle},  \label{eq:PgGSE0}
\end{equation}  
where $g^2 = \case{1}{3}(g_1^2 + g_2^2 + g_3^2)$ is the average of 
$(2 \delta \varepsilon_{\mu}/ \mu_B B)^2$ over all directions of $\vec B$
and $\langle g^2 \rangle$ is its average over the ensemble of grains.
In random matrix theory \cite{Mehta}, this distribution is known as
the Laguerre ensemble. 
Without loss of generality we may assume
that $g_1^2 \le g_2^2 \le g_3^2$. Figure \ref{fig:3} shows the
averages $\langle g_j^2 \rangle$ and a realization of the principal
$g$-factors $g_1$, $g_2$, and $g_3$ for a specific sample, as a
function of a parameter $\lambda \sim (\tau_{\rm so} \Delta)^{-1/2}$
measuring the strength of the spin-orbit scattering. (A formal
definition of $\lambda$ in a random-matrix model will be
given below.) {}From the figure, one readily observes that,
typically, the three principal $g$-factors differ by a factor
$2$--$3$.  This implies that, in spite of the average rotational
symmetry of the grains, the response of a given level
$\varepsilon_{\mu}$ to an applied magnetic field is highly
anisotropic because of mesoscopic fluctuations. 
The mathematical origin of this effect is the ``level repulsion''
factor $|g_i^2 - g_j^2|$ in the probability distribution
(\ref{eq:PgGSE0}), which signifies that, to a certain extent, ${\cal G}_{\mu}$
can be viewed a as a ``random matrix''.

Let us now turn to a more detailed discussion of our results.  Without
magnetic field, the Hamiltonian ${\cal H}$ of the grain is invariant under
time-reversal, so that all eigenstates come in doublets
$|\psi_{\mu}\rangle$ and $|{\cal T} \psi_{\mu}\rangle$, where ${\cal
T} \psi = i \sigma_2 \psi^{*}$ is the time-reversal operator. To study
the splitting of the doublets by a magnetic field, we add a term $
\mu_B \vec B \cdot \vec \sigma$ to ${\cal H}$, $\vec \sigma =
(\sigma_1,\sigma_2,\sigma_3)$ being the vector of Pauli matrices.
From degenerate perturbation theory we find that a level
$\varepsilon_{\mu}$ is split into $\varepsilon_{\mu} \pm \delta
\varepsilon_{\mu}$, with $\delta \varepsilon_{\mu}$ of the form
(\ref{eq:deltaE}). For the real symmetric $3 \times 3$ matrix
${\cal G}_{\mu}$ one has
\begin{equation}
 {\cal G}_{\mu} = G_{\mu}^{\rm T} G_{\mu},
\end{equation}
where $G_{\mu}$ is a real $3 \times 3$ matrix with elements
\begin{eqnarray}
  (G_{\mu})_{1j} + i (G_{\mu})_{2j} &=&
   - 2 \langle {\cal T} \psi_{\mu} | \sigma_j | \psi_{\mu} \rangle
  \nonumber \\
  (G_{\mu})_{3j} &=&
    2\langle \psi_{\mu} | \sigma_{j} | \psi_{\mu} \rangle,
       \label{eq:gpsi} 
\end{eqnarray}
We use random-matrix theory (RMT) to compute the distribution 
of ${\cal G}_{\mu}$. In RMT, the microscopic Hamiltonian ${\cal H}$ is
replaced by a $2N \times 2N$ random hermitian matrix $H$, where
at the end of the calculation the limit $N \to \infty$ is taken. 
(The factor $2$ accounts for spin.) 
The wavefunction $\psi_{\mu}(\vec r)$
is replaced by an $N$-component spinor eigenvector $\psi_{\mu n}$ of
$H$, where $n$ is a vector index. 
To study the effect of spin-orbit scattering, we
take $H$ of the form
\begin{mathletters} \label{eq:HSA}
\begin{equation}
  H(\lambda) = S \otimes \openone_2 + i {\lambda \over \sqrt{4N}} 
  \sum_{j} A_j \otimes \sigma_j, 
\end{equation}
where $S$ ($A_j$) is a real symmetric (antisymmetric) $N \times N$
matrix with the Gaussian distribution
\begin{eqnarray}
  P(S)   &\propto& e^{- (\pi^2/4 N \Delta^2)\, {\rm tr}\, S^{\rm T} S}, 
  \label{eq:distr}\\
  P(A_j) &\propto& e^{- (\pi^2/4 N \Delta^2)\, {\rm tr}\, A_j^{\rm T} A_j},\ 
  \ j=1,2,3. \nonumber
\end{eqnarray}
\end{mathletters}
The Hamiltonian $H(\lambda)$ is similar to the Pandey-Mehta
Hamiltonian used to describe the effect of time-reversal symmetry
breaking in a system of spinless particles \cite{Pandey}.  In Eq.\
(\ref{eq:distr}), $\Delta$ is the average spacing between the Kramers
doublets near $\varepsilon=0$. 
The amount of spin-orbit scattering is measured by the
parameter $\lambda \sim (\tau_{\rm so} \Delta)^{-1/2}$ \cite{Halperin}. 
The case $\lambda=0$ corresponds to the absence
of spin-orbit scattering, when $H = S$ is a member of the Gaussian
Orthogonal Ensemble (GOE) of random matrix theory. The case
$\lambda=(4N)^{1/2}$ corresponds to the case of strong spin-orbit
scattering, when $H$ is a member of the Gaussian Symplectic Ensemble
(GSE). The ensemble of Hamiltonians $H(\lambda)$ corresponds to a
crossover from the GOE to the GSE. Similar crossovers were studied
previously in the literature, in particular for the cases GOE--GUE
and GSE--GUE (GUE is Gaussian Unitary Ensemble) 
\cite{Pandey,French,Sommers,Falko,VanLangen}.

The distribution of the tensor ${\cal G}_{\mu}$ for an
eigenvalue $\varepsilon_{\mu}$ of the matrix $H(\lambda)$ is
related to the statistics of eigenvectors of $H(\lambda)$ in this
crossover ensemble. To deal with the twofold degeneracy of the
eigenvalue $\varepsilon_{\mu}$, we combine the two $N$-component spinor 
eigenvectors
$\psi_{\mu}$ and ${\cal T}\psi_{\mu}$ into a single $N$-component vector
of quaternions $\bar \psi = (\psi,{\cal T}\psi)$ \cite{Mehta,quaternion}.
The quaternion vector $\bar \psi$ can be parameterized as,
\begin{equation}
  \bar \psi = \sum_{k=0}^{3} \alpha_k u_k \otimes \phi_k, \label{eq:barpsiphi}
\end{equation}
where the $u_k$ are quaternion numbers with $\mbox{tr}\, u_k^{\dagger}
u_l = 2 \delta_{kl}$ (``quaternion phase factors''), the $\phi_k$ are
$N$-component real orthonormal vectors, and the $\alpha_k$ are positive
numbers such that $\sum_{k} \alpha_k^2 = 1$
($k,l=0,1,2,3$). A eigenvector in the GOE corresponds to $\alpha_0 = 1$,
$\alpha_1 = \alpha_2 = \alpha_3 = 0$, while an eigenvector in the GSE
has typically $\alpha_0 \approx \alpha_1 \approx \alpha_2 \approx \alpha_3
\approx \case{1}{2}$. A similar parameterization has been applied to the
GOE--GUE crossover \cite{French}. 
Orthogonal invariance of the distributions of $S$ and $A_j$,
together with the freedom to choose the overall quaternion phase of
$\bar \psi$, give a distribution of the $u_k$ and $\phi_k$
that is as random as possible, provided the above mentioned
orthogonality constraints are obeyed. Hence, all nontrivial information 
about the eigenvector statistics is encoded in the numbers $\alpha_k$. 
Substitution of the parameterization (\ref{eq:barpsiphi}) into Eq.\ 
(\ref{eq:gpsi}) yields
\begin{eqnarray}
%  g_{j} &=& 2(\alpha_0^2 - \alpha_1^2  - \alpha_2^2 - \alpha_3^2) +
%            4 \alpha_j^2, \ \ j=1,2,3. \label{eq:ga}
 g_{1} &=& 2(\alpha_0^2 + \alpha_1^2  - \alpha_2^2 - \alpha_3^2), \nonumber \\
 g_{2} &=& 2(\alpha_0^2 - \alpha_1^2  + \alpha_2^2 - \alpha_3^2), 
 \label{eq:ga}\\
 g_{3} &=& 2(\alpha_0^2 - \alpha_1^2  - \alpha_2^2 + \alpha_3^2). \nonumber
\end{eqnarray}
While the squares $\alpha_k^2$ ($k=0,1,2,3$) are all positive, the
principal $g$-factors as given by Eq.\ (\ref{eq:ga}) can also be
negative. Permutations of the $\alpha_k$ alter the signs of the
individual $g_j$, but not of their product $g_1 g_2 g_3$. [The product
$g_1 g_2 g_3 = \det G$ also follows from Eq.\ (\ref{eq:gpsi}); one 
verifies that it does not change when $|\psi\rangle$ is
replaced by a linear combination of $|\psi\rangle$ and $|{\cal
T}\psi\rangle$.]  Without loss of generality, we may assume that
$g_1^2 \le g_2^2 \le g_3^2$, and that $g_2$ and $g_3$ are
positive. Then equation (\ref{eq:ga}) provides the constraint
$g_2 + g_3 \le 2 + g_1$, which poses a bound on the occurrence of 
negative values for the product $g_1 g_2 g_3$.
We conclude that all information on the eigenvector statistics in the
GOE--GSE crossover is encoded in the magnitudes of $g_1$, $g_2$, and
$g_3$ and the sign of their product. Since for the
level splitting $\delta \varepsilon_{\mu}(\vec B)$ only the squares
$g_j^2$ are of relevance, we disregard the sign of $g_1 g_2 g_3$ in the
remainder of the paper. The sign of $g_1 g_2 g_3$ may be determined
in principle, however, by a spin-resonance experiment \cite{spinres}.

In order to calculate the distribution $P(g_1,g_2,g_3)$ one has, in
principle, to carry out the same program as was done in Refs.\
\onlinecite{Sommers,Falko} for the GOE--GUE crossover. However, it
turns out that in the present case the calculation is considerably
more complicated. This can already be seen from the mere observation
that the wavefunction statistics in the GOE--GSE crossover is governed by
three variables $g_1$, $g_2$, and $g_3$, whereas in the case of
the GOE--GUE crossover only one variable was needed 
\cite{Sommers,Falko,VanLangen}. In the
field-theoretic language of Ref.\ \onlinecite{Falko}, one has to use a
nonlinear sigma model of $16 \times 16$ supermatrices, instead of the
usual $8 \times 8$ for the GOE--GUE crossover \cite{Klaus}. Here we
refrain from such a truly heroic enterprise. Instead we focus on
the regimes of strong and weak spin-orbit coupling, and study the
intermediate regime by means of numerical simulations of the model
(\ref{eq:HSA}).

Before we address the case of strong spin-orbit scattering 
$\lambda \gg 1$ in the crossover Hamiltonian, we first consider
the GSE, corresponding to $\lambda^2 = 4N$. In the GSE, the
wavefunction $\psi$ is a vector of independently Gaussian distributed
complex numbers. Then, one easily verifies that, for large $N$, the
elements of the matrix $G$ of Eq.\ (\ref{eq:gpsi}) are real random
variables, independently distributed, with a Gaussian distribution of
zero mean and variance $2/N$. Hence $G$ is a random real matrix with
distribution
\begin{equation}
  P(G) \propto \exp(-N {\rm tr}\, G^{\rm T} G/4). \label{eq:GProb}
\end{equation}
The principal $g$-factors are the eigenvalues $g_{j}^2$ of the product
${\cal G} = G^{\rm T} G$. The distribution of the eigenvalues of such
a matrix product is known in literature \cite{Brezin}. It is given by
Eq.\ (\ref{eq:PgGSE0}) with $\langle g^2 \rangle = 6/N$.

\begin{figure}
\epsfxsize=0.89\hsize
%\hspace{0.1\hsize}
\epsffile{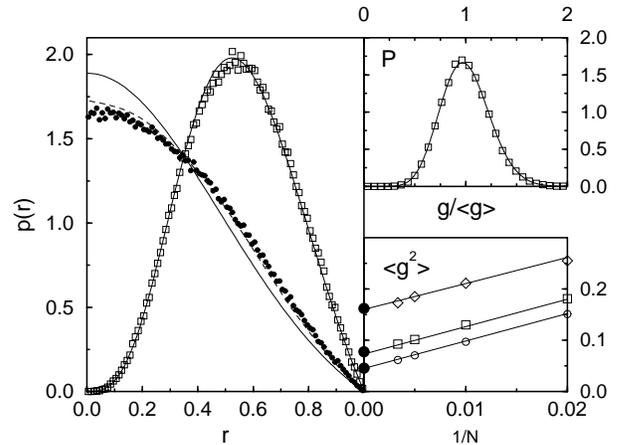}
\caption{\label{fig:1} Distribution of the orientationally averaged
$g$-factor $g^2 = (g_1^2 + g_2^2 + g_3^2)/3$ (upper left) and of the
ratios $r_{12} = |g_{1}/g_{2}|$ (circles) and $r_{23} = |g_{2}/g_{3}|$ 
(squares, main figure).  The solid curves are computed from the theory
(\protect\ref{eq:Pr12}), the data points are numerical simulations of
the random matrix model (\protect\ref{eq:HSA}) 
with $N=200$ and $\lambda = 7.7$. 
The slight discrepancy between theory and simulations for $r_{12}$
is a finite-$N$ effect; good agreement is obtained with the
GSE distribution with $N=200$ (dotted curve).
The lower inset shows $\langle g^2 \rangle$ vs. $1/N$ for
$\lambda=4.3$ (diamonds), $6.2$ (squares), and $8.1$ 
(open circles), together with
the theoretical prediction $\langle g^2 \rangle = 3\lambda^{-2}$ for 
$N \to \infty$ (closed circles).}
\end{figure}

Let us now turn to the Hamiltonian $H(\lambda)$ for large $\lambda \gg 1$,
but still $\lambda \ll N^{1/2}$. In that case, spin-rotation invariance is
broken globally (so that a wavefunction as a whole does not have a
well-defined spin), but not locally; on short length scales, the
particle keeps a well-defined spin. We then argue that, in the random
matrix language, one may think of the quaternion wavevector $\bar \psi$
as consisting of $\sim \lambda^2 \gg 1$ components,
each with a well-defined spin (or ``quaternion phase''), but with
uncorrelated spins for each component. The distribution of ${\cal G}$ is
then given by the distribution for the GSE with $N$
replaced by a number $\sim \lambda^2$ \cite{phaserigidity}.
We have found that the precise
correspondence is $N \to 2 \lambda^2$, by estimating the
exponential term in the exact distribution, along the lines of Ref.\
\onlinecite{Sommers,phaserigidity}. 
In order to verify this statement we have numerically generated
random matrices of the form (\ref{eq:HSA}). The comparison with the
GSE distribution with $N$ replaced by $2 \lambda^2$ is excellent, see
Fig.\ \ref{fig:1}.

In order to further analyze $P({\cal G})$ for strong spin-orbit scattering, 
we introduce the orientationally 
averaged $g$-factor,
\begin{equation}
  g^2 = \case{1}{3}(g_1^2 + g_2^2 + g_3^2) 
      = \left \langle { (2\delta \varepsilon_{\mu} / \mu_B |B|)^2} 
\right \rangle_{\Omega}, 
\end{equation}
where the brackets $\langle \ldots \rangle_{\Omega}$ indicate an
average over all directions of the magnetic field. Further, we
introduce the ratios $r_{12} = |g_1/g_2|$ and $r_{23} = |g_2/g_3|$ to
characterize the anisotropy of ${\cal G}$.
Changing variables in Eq.\ (\ref{eq:PgGSE0}), we find that
$P(g,r_{12},r_{23})$ reads
\begin{eqnarray}
  P 
  &\propto&  %  \nonumber
  {r_{23}^3 (1-r_{23}^2) (1-r_{23}^2 r_{12}^2)
  (1-r_{12}^2) \over (1 + r_{23}^2 + r_{23}^2 r_{12}^2)^{9/2}}\, 
  g^8 e^{-9 g^2/2 \langle g^2 \rangle}.\! \label{eq:Pr12}
\end{eqnarray}
Note that the distribution of $r_{12}$ and $r_{23}$ does not depend on
$\langle g^2 \rangle$ (provided the spin-orbit scattering is
sufficiently strong). The ``$g$-factor'' $g_z$ for a magnetic field in
the $z$-direction (which is a random direction with respect to the
principal axes) is given by $g_z = ({\cal G}_{zz})^{1/2}$. Its
distribution follows from Eq.\ (\ref{eq:GProb}) as $P(g_z) \propto
g_z^2 \exp(-3 g_z^2/2\langle g^2 \rangle)$, in agreement with Ref.\
\onlinecite{Matveev}.

\begin{figure}
\epsfxsize=0.99\hsize
%\hspace{0.1\hsize}
\epsffile{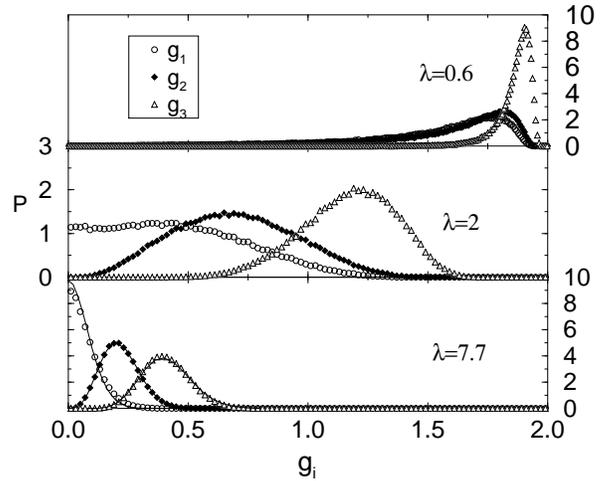}\vspace{-0.5cm}

\caption{\label{fig:2} Distributions of the principal $g$-factors
$g_1$, $g_2$, $g_3$ for $\lambda=0.6$, $2.0$, and $7.7$. 
The data points are obtained from numerical simulation of Eq.\ 
(\protect\ref{eq:HSA})
with $N=100$.} 
\end{figure}

The case of weak spin-orbit scattering can be addressed by treating
the terms proportional to $\lambda$ in Eq.\ (\ref{eq:HSA}) as a small
perturbation. To second order in $\lambda$ we find,
\begin{equation}
  {\cal G} =
  4 - {4 \lambda^2} \sum_{\nu \neq \mu} 
      a_{\mu\nu}^{\rm T} a_{\mu\nu}^{\vphantom{no superscript {\rm T}}}
  {1 \over  (\varepsilon_{\nu} - \varepsilon_{\mu})^2},
  \label{eq:pert}
\end{equation}
where $\Delta$ is the mean level spacing and $a_{\mu\nu}$ is an
antisymmetric $3 \times 3$ matrix proportional to the matrix elements
of the perturbation in the eigenbasis $\{ |\psi_{\nu} \rangle \}$ of
$H(0) = S$, $(a_{\mu\nu})_{ij} = N^{-1/2} \langle
\psi_{\mu} | A_k | \psi_{\nu} \rangle \varepsilon_{kij}$,
where $\varepsilon_{kij}$ is the antisymmetric tensor.
We first consider the change in the principal $g$-factors due to
the matrix element $a_{\mu\nu}$ coupling the level
$\varepsilon_{\mu}$ to a close neighboring level $\varepsilon_{\nu}$
where $\nu = \mu + 1$ or $\mu - 1$. (Level repulsion rules out
the possibility that both levels $\varepsilon_{\mu \pm 1}$
are very close.)  In view of the energy denominators in Eq.\
(\ref{eq:pert}), we may expect that this contribution is
dominant. Taking only the relevant matrix element $a_{\mu \nu}$ into
account, we find
\begin{equation}
  g_3 = 2,\ \ g_1 = g_2 = 2 - {\case{1}{2} \lambda^2} 
    {(\varepsilon_{\mu} - \varepsilon_{\nu})^{-2}} 
    {\rm tr}\, a_{\mu\nu}^{\rm T} 
    a_{\mu\nu}^{\vphantom{no superscript {\rm T}}},
\end{equation}
where $\nu = \mu \pm 1$. 
Since the spacing distribution $P(|\varepsilon_{\mu}
- \varepsilon_{\nu}|) \approx \pi \Delta^{-2}|\varepsilon_{\mu} -
\varepsilon_{\nu}|$ for small $\varepsilon_{\mu} - \varepsilon_{\nu}$
\cite{Mehta}, we find that the distribution $P(g)$ of both $g_1$ and
$g_2$ has tails $P(g) = (3 \lambda^2 / 2 \pi) (2 - g)^{-2}$ for $2-g
\gg \lambda^2$.  The main effect of contributions from the other
energy levels in Eq.\ (\ref{eq:pert}) is a reduction of $g_3$
below $2$, and a separation of $g_1$ and $g_2$. This is illustrated
in Fig.\ \ref{fig:3}.
The three regimes of weak, intermediate, and strong spin-orbit
scattering are compared in Fig.\ \ref{fig:2}, using a numerical 
evaluation of the distributions of the three principal $g$-values.

We gratefully acknowledge discussions with T. A. Arias, D. Davidovic,
K. M. Frahm, Y. Oreg, D. C. Ralph, and M. Tinkham. 
Upon completion of this project,
we learned of Ref.\ \onlinecite{Matveev}, 
which contains some overlap with our work.
This work was supported in part by the
NSF through the Harvard MRSEC (grant DMR 98-09363), and by grant DMR
99-81283.


\begin{references}
\bibitem{quaternion}
A quaternion is a $2 \times 2$ matrix $q$ of the form
$  
  q = q_0 \openone + i \sum_{j} q_j \sigma_j,
$
where the $q_j$ are real numbers ($j=0,1,2,3$).
\bibitem{spinres} For example, if the principal axes of ${\cal G}$ are
labeled $\hat e_1$, $\hat e_2$, and $\hat e_3 = \hat e_1 \times \hat e_2$,
and we apply a static field $B \hat e_3$, then a
resonant AC field $\vec b \propto \mbox{Re}\, [(g_2 \hat e_1 + i \eta
\hat e_2 g_1) e^{-i \omega t}]$, with $\omega = g_3 |\mu_B| B/\hbar > 0$,
will produce spin flips for $\eta=1$ but not for $\eta=-1$.
\bibitem{phaserigidity}
The same is true for the GOE--GUE crossover, where the relevant 
quantity is the ``phase-rigidity'' $\rho = |\langle {\cal T} \psi |
\psi \rangle|^2$. The distribution $P(\rho)$ for large magnetic fields 
equals $P(\rho)$ in the GUE, with $N$ replaced by $2 \alpha^2$, where
$\alpha$ is a crossover parameter analogous to $\lambda$, see Ref.\
\onlinecite{VanLangen}.
\end{references}
\end{document}